\documentclass[12pt,fleqn]{article}
\usepackage{bm}

\begin{document}

\begin{center}

{\bf \Large Unphysical Gauge Fixing in Higgs Mechanism }

\vspace{1cm}

Takehisa Fujita\footnote{e-mail:
fffujita@phys.cst.nihon-u.ac.jp}, 
Atsushi Kusaka\footnote{e-mail:
kusaka@phys.cst.nihon-u.ac.jp},
Kazuhiro Tsuda\footnote{e-mail:
nobita@phys.cst.nihon-u.ac.jp},
and Sachiko Oshima\footnote{e-mail:
oshima@phys.cst.nihon-u.ac.jp}

Department of Physics, Faculty of Science and Technology, 

Nihon University, Tokyo, Japan

\vspace{2cm}

{\Large Abstract}

\end{center}

The {\it unitary gauge} in the Higgs mechanism is to impose the condition of 
$\phi =\phi^\dagger $ on the Higgs fields. However, this is not the gauge fixing  
but simply a procedure for producing the massive vector boson fields by hand. 
The Lagrangian density of the weak interactions should be reconsidered 
by starting from the massive vector boson fields which couple to the fermion currents 
as the initial ingredients.


\section{ Introduction}
The whole idea of the symmetry breaking has been critically examined in the recent 
textbook \cite{fujita}, and the physics of the spontaneous symmetry breaking is, by now, 
understood in terms of the standard knowledge of quantum field theory.  
In particular, if one wishes to understand the vacuum state in a field theory model 
of fermions, then one has to understand the structure of the negative energy states of 
the corresponding field theory model. The importance of the negative energy state 
in the Dirac field can be easily understood if one looks into the Dirac equation. 
In the Dirac equation, the energy eigenvalue itself is the physical observable 
since it is written as
$$ \left( -i\bm{\nabla} \cdot \bm{\alpha} +m \beta \right) \psi (\bm{r},t) =
E \psi (\bm{r},t)   \eqno{(1.1)}  $$
which means that the energy eigenvalue $E$ must be physical. Therefore, if one obtains 
the energy eigenvalue which is negative, then this negative energy itself must be 
physical. For this, we have to always carry out the field quantization with 
anti-commutation relation, and this is equivalent to the realization 
of the Pauli principle. Under this Pauli principle, one can construct the vacuum 
state of the corresponding field theory model by filling out all the negative energy 
states completely. If there is any hole in the vacuum state, then this corresponds 
to the anti-particle state which is an observed fact. 

However, it is normally very difficult to construct the vacuum state of 
the interacting system, and in fact, the exact solution of the model field 
theory is practically impossible in four dimensions. Nevertheless, the physics 
of the spontaneous symmetry breaking is now clearly understood and some 
of the model field theory prefer the vacuum state which breaks the chiral symmetry 
though there exists no massless (Goldstone) boson. 
In this respect, one can say that the symmetry breaking physics can be understood 
only after one solves the whole system of the corresponding field theory model. 
One cannot understand its physics by rewriting the Lagrangian density into 
a new shape since the Lagrangian density itself cannot be any physical observables. 
From the Lagrangian density, one can learn a symmetry property of the model 
field theory. 

In this respect, the physics of the Higgs mechanism \cite{higgs} is very vague, and 
as we will show below, the whole procedure of the Higgs mechanism was carried out 
with simple-minded mistakes. This is mainly connected to the misunderstanding 
of the gauge fixing where one degree of freedom of the gauge fields must be reduced 
in order to solve the equations of motion of the gauge fields. Therefore, one cannot 
insert the condition of the gauge fixing into the Lagrangian density. This is clear since 
the Lagrangian density only plays a role for producing the equation of motions.  
Indeed, the Lagrangian density itself is not directly a physical observable, and 
the Hamiltonian constructed from the Lagrangian density is most important after 
the fields are quantized. For the field quantization, one has to make use of the gauge 
fixing condition which can determine the gauge field $A_\mu$ together with the equation 
of motions. This means that 
only the final Lagrangian density is relevant to the description of physical observables, 
and thus the success of the Glashow-Weinberg-Salem model \cite{glashow,weinberg,salam} is 
entirely due to the final version of the weak Hamiltonian which is not at all 
the gauge field theory but is a model field theory of the massive vector fields 
which couple to the fermion currents.

\vspace{0.5cm}
\section{ Gauge Fixing }

The Lagrangian density of the Higgs mechanism is given as
$$  {\cal L} = {1\over 2}  (D_\mu\phi)^\dagger (D^\mu\phi) -{1\over 4}
u_0 \left(  |{\phi}|^2
-\lambda^2 \right)^2 -{1\over 4} F_{\mu\nu} F^{\mu\nu}  \eqno{(2.1)}   $$
where
$$  D_{\mu}= {\partial}_{\mu} + ig A_{\mu}, \quad
  F_{\mu\nu}= \partial_{\mu} A_{\nu} - \partial_{\nu} A_{\mu} . \eqno{(2.2)} $$
Here, we only consider the U(1) case since it is sufficient for the present 
discussions. The above Lagrangian density is indeed gauge invariant, and 
in this respect, the scalar field may interact with gauge fields in eq.(2.1). 
However, it should be noted that there is no experimental indication that 
the fundamental scalar field can interact with any gauge fields in terms of 
the Lagrangian density of eq.(2.1). 

The equations of motion for the scalar field $\phi$ become 
$$ \partial_\mu ({\partial}^{\mu} + ig A^{\mu})\phi = -u_0 \phi \left(  |{\phi}|^2
-\lambda^2 \right) - ig A_\mu ({\partial}^{\mu} + ig A^{\mu}) \phi   \eqno{(2.3)}  $$
$$ \partial_\mu ({\partial}^{\mu} - ig A^{\mu})\phi^\dagger = -u_0 \phi^\dagger 
\left(  |{\phi}|^2-\lambda^2 \right) + ig A_\mu ({\partial}^{\mu} - ig A^{\mu}) 
\phi^\dagger . \eqno{(2.4)}    $$
On the other hand, the equation of motion for the gauge field $A_{\mu}$ can be 
written as 
$$ \partial_\mu F^{\mu \nu} = gJ^\nu  \eqno{(2.5)} $$
where
$$ J_\mu ={1\over 2}i \left\{ \phi^\dagger ({\partial}_{\mu} + ig A_{\mu}) \phi 
-\phi({\partial}_{\mu} - ig A_{\mu}) \phi^\dagger  \right\}. \eqno{(2.6)} $$ 
One can also check that the current $ J_\mu$ is conserved, that is
$$ \partial_\mu  J^\mu =0. \eqno{(2.7)} $$
This Lagrangian density of eq.(2.1) 
has been employed for the discussion of the Higgs mechanism. 

\subsection{Gauge Freedom and Number of Independent Equations}
Now, we should count the number of the degrees of freedom and the number of 
equations. For the scalar field, we have two independent functions $\phi$ and 
$\phi^\dagger$. Concerning the gauge fields $A_{\mu}$, we have four since there are 
$A_0$, $A_1$, $A_2$, $A_3$ fields. Thus, the number of the independent fields 
is six. On the other hand, the number of equation is five since 
the equation for the scalar fields is two and the number of the gauge fields 
is three. This three can be easily understood, even though it looks that 
the independent number of equations in eq.(2.5) is four, but due to the current 
conservation the number of the independent equations  becomes three. 
This means that the number of the independent 
functions is six while the number of equations is five, and they are not equal. 
This is the gauge freedom, and therefore in order to solve the equations of 
motion, one has to put an additional condition for the gauge field  $A_{\mu}$ 
like the Coulomb gauge which means $ \bm{\nabla } \cdot \bm{A} =0 $.    
In this respect, the gauge fixing is simply to reduce the redundant 
functional variable of the gauge field $A_{\mu}$ to solve the equations of motion, 
and nothing more than that.

\subsection{Unitary Gauge Fixing}
In the Higgs mechanism, the central role is played by the gauge fixing 
of the unitary gauge. The unitary gauge means that one takes 
$$ \phi=\phi^\dagger . \eqno{(2.8)} $$
This is the constraint on the scalar field $\phi$ even though there is no 
gauge freedom in this respect. For the scalar field, the phase can be changed, 
but this does not mean that one can erase one degree of freedom. 
One should transform the scalar field in the gauge transformation as
$$ \phi' = e^{-ig \chi} \phi $$
but one must keep the number of degree of freedom after the gauge transformation. 
Whatever one fixes the gauge $\chi$, one cannot change the shape of the scalar field 
$\phi$ since it is a functional variable and must be determined from the equations 
of motion. The gauge freedom is, of course, found in the vector potential $A_{\mu}$  as 
we discussed above. In this sense, one sees that the unitary gauge fixing is a simple 
mistake. The basic reason why people overlooked 
this simple-minded mistake must be due to their obscure presentation of the Higgs 
mechanism. Also, it should be related to the fact that, at the time of presenting 
the Higgs mechanism, the spontaneous symmetry breaking physics was not understood 
properly since the vacuum of the corresponding field theory was far beyond the proper 
understanding. Indeed, the Goldstone boson after the spontaneous symmetry breaking 
was taken to be almost a mysterious object since there was no experiment which suggests 
any existence of the Godlstone boson. Instead, a wrong theory prevailed among physicists. 
Therefore, they could assume a very unphysical 
procedure of the Higgs mechanism and people pretended that they could understand it all. 
 
\subsection{Final Lagrangian Density}

After an improper gauge fixing, one arrives at the final Lagrangian density
$$  {\cal L} = {1\over 2}  (\partial_\mu\eta) (\partial^\mu\eta) -{1\over 4}
u_0 \left(  |\lambda + \eta(x)|^2 -\lambda^2 \right)^2 
 +{1\over 2}g^2(\lambda+ \eta(x))^2 A_\mu A^\mu -{1\over 4} F_{\mu\nu} F^{\mu\nu}
 \eqno{ (2.9)} $$
where we rewrite the Higgs field as
$$ \phi =\phi^\dagger =\lambda + \eta(x). \eqno{ (2.10)} $$
Since the real scalar field $\eta $ is supposed to be small and besides a real 
scalar field is unphysical \cite{fkks,kkof}, it may be set to zero, that is, 
$\eta=0$. In this case, we arrive at the following Lagrangian density 
$$  {\cal L} =  {1\over 2}g^2\lambda^2 A_\mu A^\mu -{1\over 4} F_{\mu\nu} F^{\mu\nu}. 
 \eqno{ (2.11)} $$
This should be the final Lagrangian density of the Higgs theory, and it is nothing 
but the massive vector boson field which has nothing to do with the gauge theory.  

\subsection{Proper Gauge Fixing}

For the equations of motion eqs.(2.3-5), one can make a proper gauge fixing such as 
the Coulomb gauge ($ \bm{\nabla} \cdot \bm{A} =0$) or the temporal gauge ($ A_0=0$). 
In this case, one can solve the equations of motion properly, and one can calculate 
any physical observables which are, in this case, gauge invariant quantities 
like $ F^{\mu \nu}$ in the classical field theory. Further, we can quantize 
the fields $\phi$ and $A^\mu$ after the gauge fixing, and we can carry out 
the perturbation theory since we have now the quantized Hamiltonian. 
But this must be completely different from the Higgs mechanism.   

\subsection{Higgs Potential}

In the Lagrangian density of the Higgs mechanism, one assumes a self-interacting 
field potential 
$$  U(\phi)= {1\over 4}u_0 \left(  |{\phi}|^2 -\lambda^2 \right)^2.   \eqno{(2.12)}   $$
This was originally introduced in the discussion of the spontaneous symmetry breaking 
physics as a toy model \cite{gold}. But in the mean time, this scalar field 
potential is considered to be a fundamental potential. However, one may ask a question 
as to where this potential $  U(\phi)$ is produced  from since the scalar field is 
obviously not a free field. Unless one can understand the basic origin of this potential 
$  U(\phi)$, it is extremely difficult to understand the Higgs mechanism itself 
from the fundamental physics point of view. 

\section{What shall we do ?}
When experiments suggested that there might be heavy vector bosons exchanged 
between leptons and baryons in the weak processes, 
people wanted to start from the massive vector bosons. However, it was somehow 
believed among educated physicists that only gauge field theories must be renormalizable. 
We do not know where this belief came from. In fact, there is no strong reason that only 
the gauge field theory is renormalizable. It is clear that QED is a good gauge 
theory which is renormalizable, and there is no conceptual problem in QED which can 
describe all the experiments related to the electromagnetic processes. However, this 
does not mean that other non-gauge field theory models are not renormalizable. In fact, 
the basic condition of the renormalizability must be concerned with the coupling constant 
$g$ which must be dimensionless \cite{fujita}.

\subsection{Renormalizability of Non-abelian Gauge Field}
However, one should be careful for the renormalizability of the non-abelian gauge 
field theory.  As one can easily convince oneself, the non-abelian gauge theory 
has an intrinsic problem of the perturbation theory \cite{fujita2}. 
This is connected to the fact 
that the color charge in the non-abelian gauge field depends on the gauge, and 
therefore it cannot be physical observables. This means that the free gauge field 
which has a color charge is gauge dependent, and thus one cannot develop 
the perturbation theory in a normal way. In QCD, this is exhibited as the experimental 
fact that free quarks and free gluons are not observed in nature. No free field is 
a kinematical constraint and thus it is beyond any dynamics. 
Therefore, one cannot discuss the renormalizability 
of the non-abelian gauge field theory models due to the lack of the perturbation 
scheme in this model field theory \cite{fujita,fujita2}. 

\subsection{Massive Vector Field Theory}

Even though the Higgs mechanism itself is meaningless, the final Lagrangian density 
may well be physically interesting. This is clear since, from this Lagrangian density, 
one can construct the Hamiltonian which can describe the experimental observables. 
In this respect, we may write the simplest Lagrangian density for two flavor leptons 
which couple to the SU(2) vector fields $W_\mu^a$ 
$$  {\cal L} = \bar{\Psi} (i\partial_\mu \gamma^\mu +m)\Psi - gJ_\mu^a W^{\mu,a}  
 +{1\over 2}M^2 W_\mu^a W^{\mu,a} -{1\over 4} G_{\mu\nu}^a G^{\mu\nu,a}  \eqno{ (3.1)} $$
where $M$ denotes the mass of the vector boson. 
The fermion wave function $\Psi$ has two components.   
The fermion current $J_\mu^a $ and the field strength $ G_{\mu\nu}^a$ are defined as 
$$ J_\mu^a =\bar{\Psi}  \gamma_\mu \tau^a \Psi, \qquad  
G_{\mu\nu}^a= \partial_{\mu} W_{\nu}^a - \partial_{\nu} W_{\mu}^a . \eqno{(3.4)} $$
This Lagrangian density is almost the same as the standard model Lagrangian density, 
apart from the Higgs fields. 

\subsection{Renormalizability}

The most important of all is to examine the renormalizability of the massive 
vector boson fields which couple to the fermion currents \cite{nishi}. 
In this sense, this is just similar to checking 
the final Lagrangian density of the standard model itself whether it can 
be renormalizable or not. This is beyond the scope of the present paper, but it should be 
done in future. 


\

\end{document}